\documentclass[12pt,twoside]{amsart}

\usepackage{amssymb}
\usepackage[mathscr]{eucal}

\newtheorem{prop}{Proposition}
\theoremstyle{definition}

\newcommand{\ds}{\displaystyle}

\def\EXP{\mbox{{\large\bf e}}}
\newcommand{\uop}{\mathbf{u}}
\newcommand{\wop}{\mathbf{w}}
\newcommand{\xop}{\mathbf{x}}
\newcommand{\zop}{\mathbf{z}}

\newcommand{\R}{\mathbf{R}}
\newcommand{\Q}{\mathbf{Q}}

\newcommand{\ltoda}{\ell}
\newcommand{\ldst}{{\widetilde{\ell}}}
\newcommand{\montoda}{\widehat{\mathbf{t}}}
\newcommand{\tp}{\mathbf{t}}
\newcommand{\Ltoda}{L}
\newcommand{\Ldst}{\widetilde{L}}
\newcommand{\Montoda}{\widehat{T}}

\newcommand{\Om}{\Delta}

\newcommand{\omeg}{{\omega^{1/2}}}
\newcommand{\omegg}{{\omega^{-1/2}}}
\newcommand{\ff}{\phi}

\newcommand{\Set}{\mathfrak{F}_M}

\begin{document}

\begin{flushright}
ITEP-TH-39/01
\end{flushright}

\vspace{2cm}

\title[]{Relativistic Toda chain at root of unity II.\\
Modified $Q$-operator}%
\author{S. Pakuliak}
\address{Stanislav Pakuliak, Bogoliubov Laboratory of Theoretical Physics, Joint Institute for Nuclear Research,
Dubna 141980, Russia and Max-Planck Institut
f\"ur Mathematik, Vivatsgasse 7, 53111 Bonn, Germany}%
\email{pakuliak@thsun1.jinr.ru}%
\author{S. Sergeev}%
\address{Sergei Sergeev, Bogoliubov Laboratory of Theoretical Physics, Joint Institute for Nuclear Research,
Dubna 141980, Russia}
\email{sergeev@thsun1.jinr.ru}

\thanks{This work was supported in part by the grant INTAS OPEN 00-00055.
S.P.'s work was supported by the grant RFBR 00-02-16477 and grant
for support of scientific schools RFBR 00-15-96557, S.S's work
was supported by the grant RFBR 01-01-00201.}%


\begin{abstract}
Matrix elements of quantum intertwiner as well as the modified
Q-operator for the quantum relativistic Toda chain at root of
unity are constructed explicitly. Modified Q-operators make
isospectrality transformations of quantum transfer matrices so
that the classical counterparts of Q-operators correspond to the
B\"acklund isospectrality transformations of the classical
transfer matrices. Separated vectors for the Functional Bethe
Ansatz are constructed with the help of modified Q-operators.
\end{abstract}

\maketitle


\tableofcontents

\section*{Introduction}

This paper is the continuation of Ref. \cite{sp-1} where the
quantum relativistic Toda chain at a root of unity was
investigated. In the previous paper we have established several
features of the chain, concerning the relationship between its
finite dimensional (spin) counterpart and its nontrivial
classical counterpart.

For several integrable models, based on the simple Weyl algebra
at $N$-th root of unity, the $\mathbb{C}$-numerical $N$-th powers
of Weyl generators form a classical discrete integrable system.
So the parameters of the unitary representation of Weyl algebra
form the classical counterpart of the finite dimensional
integrable system (i.e. spin integrable system), see
\cite{FK-qd,BR-qd,BR-unpublished,Sergeev}. Most important finite
dimensional operators, arising in the spin systems, have
functional counterparts, defined as rational mappings in the
space of $N$th powers of the parameters. So the finite dimensional
operators are the secondary objects: one has to define first the
mapping of the parameters, and then the finite dimensional
operators are to be constructed in the terms of initial and final
values of the parameters. Usual finite dimensional integrable
models correspond to the case when for all the operators involved
the initial and final parameters coincide. Algebraically these
conditions are the origin e.g. of Baxter's curve in the case of
the chiral Potts model \cite{CPM,BS-cpm}, or of the spherical
triangle parameterization for the Zamolodchikov-Bazhanov-Baxter
model.

The simplest case of such combined system is the so-called
quantum Toda chain at root of unity. It differs significantly
from the usual Toda chain \cite{Sklyanin-sep} as well as from the
relativistic Toda chain with arbitrary $q$ \cite{PS}. For RTC at
root of unity $L$-operators, quantum as well as classical, and
the quantum intertwiners have rather simple form \cite{PS}. In
\cite{sp-1} it was shown that the trace of monodromy of quantum
intertwiners, giving usually Baxter's $Q$-operator, has the
nontrivial classical counterpart. Under several special
additional conditions, the classical counterpart of $Q$ operator
is the B\"acklund transformation \cite{Sklyanin-rev,KS-manybody}
for the classical counterpart of relativistic Toda chain. Spin
counterpart of the corresponding B\"acklund transformation makes
an isospectral transformation of the spin system. Parameters of
Weyl algebra are described in the terms of
$\tau$-function\footnote{in this paper we use the term
``$\tau$-function'' in the sense of bilinear discrete equation
with constant coefficients, i.e. as a trigonometric limit of the
$\theta$-function on a high genus algebraic curve}, and the
transformation mentioned maps $(n-1)$-solitonic state into
$n$-solitonic state. Initial homogeneous state is supposed to be
$0$-solitonic, while the maximal number of solitons is $M-1$ for
the chain of the length $M$. The isospectral transformation of
the spin system depends on the $(n-1)$-solitonic origin and
$n$-solitonic image of the classical states. Finite dimensional
similarity operators, called the modified $Q$-operators, of
course, do not form a commutative family. Nevertheless these
isospectrality deformations play the important r\^ole in the
Functional Bethe Ansatz \cite{sp-1} (see
\cite{Sklyanin-fun,Sklyanin-sep,Lebedev} for the details
concerning the method of FBA). Namely, for a special set of the
solitonic amplitudes the similarity operator becomes a projector
into an eigenstate of non-diagonal element of the quantum Toda
chain monodromy matrix. It was shown in \cite{sp-1}, but we did
not give explicit formulas for matrix elements of all finite
dimensional operators (quantum intertwiner, similarity operators
as well as the separating operator). All these is the subject of
the current paper.

This paper is organized as follows. First, we recall the
formulation of the quantum relativistic Toda chain at root of
unity and its classical counterpart. Next, we recall the form of
the auxiliary $L$-operators (which are the particular case of
Bazhanov-Stroganov's $L$-operator \cite{BS-cpm}), its classical
counterpart, the intertwining relations and the isospectrality
transformations of the Toda transfer matrix. In the third section
the matrix elements of quantum $R$-matrix and modified
$Q$-operator are given in an appropriate basis. In the fourth
section we recall the parameterization of $\tau$-function in
application to the relativistic Toda chain. In the fifth section
we parameterize arguments of the modified $Q$-operators for the
whole set of B\"acklund transformations and so construct
explicitly the similarity operator transforming the homogeneous
chain into the general $(M-1)$-parametric inhomogeneous chain. In
the sixth section, we describe an appropriate limiting procedure
for the degenerative final state and construct the quantum
separating operator. Finally, in conclusion, we discuss several
further applications of the method described.

\section{Formulation of the model}

In this section we recall briefly the subject of the model called
the quantum relativistic Toda chain at root of unity
\cite{sp-1,PS}.

\subsection{$L$-operators}

Let the chain is formed by $M$ sites with the periodical boundary
conditions. $m$-th site of the Toda chain is described by the
following local Lax matrix:
\begin{equation}\label{l-toda}
\ds\ltoda_m(\lambda)\;=\;\left(\begin{array}{rcr} \ds
1\,+\,{\kappa\over\lambda}\,\uop_m^{}\,\wop_m^{} &,&
\ds -\,{\omeg\over\lambda}\,\uop_m^{} \\
&&\\
\ds \wop_m^{} &,& 0 \end{array}\right)\;.
\end{equation}
Here $\lambda$ is the spectral parameter,  $\kappa$ is an extra
complex parameter, common for all sites, i.e. the modulus.
Elements $\uop_m$ and $\wop_m$ form the ultra-local Weyl algebra,
\begin{equation}\label{Weyl-algebra}
\ds \uop_m^{}\cdot\wop_m^{}\,=\,
\omega\;\wop_m^{}\cdot\uop_m^{}\;,
\end{equation}
and $\uop_m$, $\wop_m$ for different sites commute. Weyl's
parameter $\omega$ is the primitive root of unity,
\begin{equation}
\ds\omega\;=\;\EXP^{2\,\pi\,i/ N}\;,\;\;\;
\omeg\;=\;\EXP^{i\,\pi/N}\;.
\end{equation}
$N$-th powers of the Weyl elements are centers of the algebra. We
will deal with the finite dimensional unitary representation of
the Weyl algebra, i.e.
\begin{equation}\label{uw-XZ}
\ds \uop_m\;=\;u_m\,\xop_m\;,\;\;
\wop_m\;=\;w_m\,\zop_m\;,
\end{equation}
where $u_m$ and $w_m$ are $\mathbb{C}$-numbers, and
\begin{equation}
\ds\xop_m = 1\otimes 1\otimes ... \otimes
\underbrace{\xop}_{m-\textrm{th}\atop\textrm{place}}
\otimes ...\,,
\;\;\;
\zop_m = 1\otimes 1\otimes ... \otimes
\underbrace{\zop}_{m-\textrm{th}\atop\textrm{place}}
\otimes ...
\end{equation}
Convenient representation of $\xop$ and $\zop$ on the
$N$-dimensional vector space $|\alpha\rangle=|\alpha\mod N\rangle$
is
\begin{equation}\label{representation}
\ds \xop\;|\alpha\rangle\;=\;|\alpha\rangle\;\omega^\alpha
\;,\;\;\;
\zop\;|\alpha\rangle\;=\;|\alpha+1\,\rangle\;,\;\;\;
\langle\alpha|\beta\rangle\;=\;\delta_{\alpha,\beta}\;.
\end{equation}
Thus $\xop$ and $\zop$ are normalized to the unity
($\xop^N=\zop^N=1$) $N\times N$ dimensional matrices, and the
$N$th powers of the local Weyl elements are $\mathbb{C}$-numbers
\begin{equation}\label{uw-N-powers}
\ds \uop_m^N\;=\;u_m^N\;,\;\;\; \wop_m^N\;=\;w_m^N\;.
\end{equation}
In general, all $u_m$ and $w_m$ are different, so we deal with
the inhomogeneous chain.

Variables $u_m^N$ and $w_m^N$ form the classical counterpart of
the quantum relativistic Toda chain, and the classical Lax matrix
is
\begin{equation}\label{L-toda}
\ds\Ltoda_m(\lambda^N)\,\stackrel{def}{=}\,
\left(\begin{array}{rcr} \ds
1\,+\,{\kappa^N\over\lambda^N}\,u_m^N w_m^N &, & \ds
{u_m^N\over\lambda^N}\\
&&\\
\ds w_m^N &,& 0
\end{array}\right)\;.
\end{equation}

\subsection{Transfer matrix and integrability}

Ordered product of the quantum $L$-operators (\ref{l-toda})
\begin{equation}\label{l-monodromy}
\ds\montoda(\lambda)\;\stackrel{def}{=}\;
\ltoda_1(\lambda)\;\ltoda_2(\lambda)\;\cdots\;
\ltoda_M(\lambda)\;=\;
\left(\begin{array}{cc} \ds
\mathbf{a}(\lambda) & \ds
\mathbf{b}(\lambda)\\
&
\\ \ds \mathbf{c}(\lambda) &
\mathbf{d}(\lambda)\end{array}\right)\;,
\end{equation}
and its trace
\begin{equation}\label{transfer}
\ds \tp(\lambda)\;=\;\mathbf{a}(\lambda)+\mathbf{d}(\lambda) \;=\;
\sum_{k=0}^{M}\; \lambda^{-k}\;\tp_k\;.
\end{equation}
are the monodromy and the transfer matrices of the model.

Transfer matrices, defined with the given $\kappa$ and with the
given set of $u_m,w_m$, $m=1...M$,
\begin{equation}
\ds\tp(\lambda)\;=\;
\tp(\lambda,\kappa\,;\{u_m,w_m\}_{m=1}^M)\;.
\end{equation}
form the commutative set: $\forall\;\;\lambda,\mu$
\begin{equation}\label{comm}
\ds \Bigl[\;\tp(\lambda,\kappa\,;\{u_m,w_m\})\;,\;
\tp(\mu,\kappa\,;\{u_m,w_m\})\;\Bigr]\;=\;0\;.
\end{equation}
Eq. (\ref{comm}) is provided by the intertwining
relation for $L$-operators (\ref{l-toda}) in the
auxiliary two-dimensional vector spaces with the help of
well known six-vertex trigonometric $R$-matrix,
see \cite{sp-1} for the details.

In the spectral decomposition of the transfer matrix
(\ref{transfer}) the utmost operators are
\begin{equation}\label{utmost}
\ds \tp_0=1\;,\;\;\;
\tp_M=\kappa^M\,\prod_{m=1}^M\,(-\omega^{1/2}u_mw_m)\,
\mathbf{Y}\;,
\end{equation}
where
\begin{equation}\label{Y}
\ds\mathbf{Y} =
\prod_{m=1}^M\,(-\omega^{-1/2}\xop_m\zop_m)\;,\;\;\;
\mathbf{Y}^N=1\;.
\end{equation}

Further we will need also the monodromy of the classical
Lax matrices (\ref{L-toda}),
\begin{equation}\label{Mon-toda}
\ds\Montoda\;=\;\Ltoda_1\;\Ltoda_2\;\cdots\;\Ltoda_M\;=\;
\left(\begin{array}{cc} \ds A(\lambda^N) & \ds
B(\lambda^N)\\
&\\
\ds C(\lambda^N) & \ds D(\lambda^N)
\end{array}\right)
\end{equation}
and its trace
\begin{equation}
\ds T(\lambda^N)=A(\lambda^N)+D(\lambda^N)\;.
\end{equation}

$M$ nontrivial coefficients in the decomposition of
$T(\lambda^N)$ are involutive with respect to the Poisson
brackets, associated with the following symplectic form:
\begin{equation}\label{symplectic}
\ds\Omega=\sum_{m=1}^M \; {d\,u_m\over
u_m}\,\wedge\,{d\,w_m\over
w_m}\;.
\end{equation}
It is the consequence of (\ref{Weyl-algebra}) and (\ref{comm}).

The utmost coefficient in the decomposition of the classical
transfer matrix  $T(\lambda^N)$, analogous to (\ref{Y}),
generates the simple gauge transformation of $L$-operators
(\ref{l-toda}). Thus we may fix without lost of generality
\begin{equation}\label{gauge-fix}
\ds \prod_{m=1}^M\,(-\omega^{1/2}u_m) \;=\;
\prod_{m=1}^M(-w_m)
\;=\;1 \;.
\end{equation}

\subsection{Quantum separation of variables}

Eigenstates of off-diagonal element of the monodromy matrix
(\ref{l-monodromy}) $\mathbf{b}(\lambda)$ play the important
r\^ole in the constructing of the spectrum of the quantum transfer
matrix. This method is known as the method of the Functional Bethe
Ansatz, or the method of the quantum separation of variables
\cite{Sklyanin-fun,Sklyanin-sep,Lebedev}. It is useful to
parameterize the spectrum of $\mathbf{b}(\lambda)$ by its zeros
$\lambda_k$, $k=1...M-1$. To deal with the eigenstates of
$\mathbf{b}(\lambda)$ explicitly is not useful. We should better
define the set of vectors $\ds
|\{\lambda_k\}_{k=1}^{M-1},\gamma\rangle$, such that
\begin{equation}\label{b-eigen}
\ds
\mathbf{b}(\lambda_j)\,|\{\lambda_k\}_{k=1}^{M-1},\gamma\rangle
\,=\,0\;,\;\;\;\lambda_j\in\{\lambda_k\}_{k=1}^{M-1}\;,
\end{equation}
and
\begin{equation}
\ds \mathbf{Y}\,|\{\lambda_k\}_{k=1}^{M-1},\gamma\rangle\,=\,
|\{\lambda_k\}_{k=1}^{M-1},\gamma\rangle\,\omega^\gamma\;,
\end{equation}
where the integral of motion $\mathbf{Y}$ is given by (\ref{Y}).

The matrix element between an eigenstate of $\tp_k$ and
$|\{\lambda\},\gamma\rangle$ is a product of $(M-1)$ Baxter's
functions $Q(\lambda_k)$, taken for the same $\lambda_k$,
$k=1...M-1$, as in (\ref{b-eigen}), see e.g.
\cite{Sklyanin-fun,Lebedev}. For the details concerning the Baxter
equation for the quantum relativistic Toda chain at root of unity
see \cite{sp-1}, and later we will recall it.  Note here, this
$Q$ is a meromorphic function on Baxter's curve. In our
parameterization it is given by
\begin{equation}\label{curve}
\ds\lambda^N\;=\;\delta^N\,+\,{\lambda^N\over\delta^N}\,+\,\kappa^N\;.
\end{equation}
Convenient Baxter's form $x^N+y^N=1+k^2x^Ny^N$ may be obtained by
the substitution $x=1/\delta$, $y=\delta/\lambda$,
$k^2=-\kappa^N$.

Spectra of $\mathbf{a}$, ..., $\mathbf{d}$ are described by $A$,
..., $D$, see \cite{Tarasov-cyclic}. For example,
\begin{equation}\label{b-spectrum}
\ds\prod_{n\in\mathbb{Z}_N}
\;\mathbf{b}(\omega^n\lambda)\;=\;B(\lambda^N)\;.
\end{equation}
Eq. (\ref{b-spectrum}) allows one to find the set of $\lambda_k$.
We will do it later for the homogeneous model.

The real reason why we are considering the classical counterpart
is that the quantum separation of variables is associated with
the B\"acklund transformation of the classical model.

\section{Intertwining relations}

In this section we recall the auxiliary $L$-operators, quantum
$R$-matrix and the Darboux relation for the classical
counterpart, providing the solution of the isospectrality
problem, construction of modified $Q$-operators and B\"acklund
transformation.

\subsection{Auxiliary $L$-operator}

Toda $L$-operators (\ref{l-toda}) are intertwined in the quantum
spaces with the Bazhanov-Stroganov $L$-operators \cite{BS-cpm}.
This fact is the origin of the relationship between the
relativistic Toda chain at the $N$th root of unity and the
$N$-state chiral Potts model. In general the intertwiner is
rather complicated, but there exists a special limit of
Bazhanov-Stroganov's $L$-operator such that the intertwiner
simplifies. This limit is the analogue of the so-called dimer
self-trapping $L$-operators for usual Toda chain, see
\cite{Sklyanin-rev,Sklyanin-sep,PS} for the details. We define
the auxiliary $L$-operator as follows:
\begin{equation}\label{ldst-uwkappa}
\ds\begin{array}{l} \ds
\ldst^{}_\ff(\lambda,\lambda_\ff)\;=\\
\\
\ds =\;\left(\begin{array}{rcr} \ds
1\,-\,\omeg\kappa_\ff^{}{\lambda_\ff^{}\over\lambda} \wop_\ff^{}
&,& \ds
-{\omeg\over\lambda}(1-\omeg\kappa_\ff^{}\wop_\ff^{})\uop_\ff^{}
\\
&&\\
\ds -\omeg\lambda_\ff^{}\uop_\ff^{-1}\wop_\ff^{} &,& \ds
\wop_\ff^{}
\end{array}\right)\;.\end{array}
\end{equation}
Here $\lambda$ and $\lambda_\ff$ are two spectral parameters
(actually, up to a gauge transformation, $\ldst$ depends on their
ratio). $\kappa_\ff$ is the module as well as $\kappa$.
$\uop_\ff$ and $\wop_\ff$ form the same Weyl algebra at the root
of unity (\ref{uw-XZ}),
\begin{equation}
\ds\uop_{\ff}\;=\;u_\ff\;\xop_\ff\;,\;\;\;
\wop_\ff\;=\;w_\ff\;\zop_\ff\;.
\end{equation}
In all these notations the subscript $\ff$ stands as the
``number'' of additional Weyl algebra.

Classical counterpart of $\ldst$ is
\begin{equation}\label{L-dst}
\ds \Ldst_\ff\;=\;\left(\begin{array}{rcr} \ds
1+\kappa_\ff^{N}{\lambda_\ff^N\over\lambda^N}w_\ff^{N} &,&
\ds{u_\ff^N\over\lambda^N}(1+\kappa_\ff^N w_\ff^N)\\
&&\\
\ds \lambda_\ff^N{w_\ff^N\over u_\ff^N} &,& w_\ff^N
\end{array}\right)\;.
\end{equation}

$L$-operators (\ref{ldst-uwkappa}) are intertwined in their two
dimensional auxiliary vector spaces by the same six-vertex
trigonometric $R$-matrix as (\ref{l-toda}). Also there exists the
fundamental quantum intertwiner for (\ref{ldst-uwkappa}): it is
the $R$-matrix for the chiral Potts model such that two
rapidities are fixed to a special singular value. The method
describing in this paper may be applied directly to the model,
defined by (\ref{ldst-uwkappa}), this is the subject of the
forthcoming paper.

\subsection{Intertwining}

We are going to write out some quantum intertwining relation for
$L$-operators (\ref{l-toda}) as well as for the whole monodromy
matrix (\ref{l-monodromy}). So we use notations, applicable for
the recursion in $m$. Also in this section we will point out
parameters $u_m,w_m$, (\ref{uw-XZ}), as the arguments of
$L$-operators.

\begin{prop}\label{prop-intertwining}
There exists unique (up to a constant multiplier) $N^2\times N^2$
matrix $\R_{m,\ff}(\lambda_\ff)$, such that $\R_{m,\ff}$,
$\ltoda_m$ and $\ldst$ obey the modified intertwining relation
\begin{equation}\label{op-recursion}
\ds\begin{array}{l}
\ds\ldst_\ff^{}(\lambda,\lambda_\ff;u_{\ff,m}^{},w_{\ff,m}^{})\cdot
\ltoda_m^{}(\lambda;u_m^{},w_m^{})\;\R_{m,\ff}^{}(\lambda_\ff)\;=\\
\\
\ds =\;
\R_{m,\ff}^{}(\lambda_\ff)\;\ltoda_m^{}(\lambda;u_m',w_m')\cdot
\ldst_\ff^{}(\lambda,\lambda_\ff;u_{\ff,m+1}^{},w_{\ff,m+1}^{})\;,
\end{array}
\end{equation}
if and only if their classical counterparts of $\Ltoda_m$ and
$\Ldst$ obey the following Darboux relation:
\begin{equation}\label{LL-darboux}
\ds\Ldst(u_{\ff,m}^{N},w_{\ff,m}^{N})
\Ltoda(u_m^{N},w_m^{N})\;=\;
\Ltoda(u_m^{\prime N},w_m^{\prime N})
\Ldst(u_{\ff,m+1}^{N},w_{\ff,m+1}^{N})\;,
\end{equation}
\end{prop}
Proof: the direct verification. Eq. (\ref{op-recursion}) may be
rewritten as
\begin{equation}\label{fundamental}
\ds \ldst_\ff^{}(\lambda;\uop_\ff^{},\wop_\ff^{})\cdot
\ltoda_m^{}(\lambda;\uop_m^{},\wop_m^{})\;=\;
\ltoda_m^{}(\lambda;\uop_m',\wop_m')\cdot
\ldst_\ff^{}(\lambda;\uop_\ff',\wop_\ff')\;,
\end{equation}
where
\begin{equation}
\ds\uop_m'=u_m'\,\R_{m,\ff}^{}\xop_m^{}\R_{m,\ff}^{-1}\;,\;\;\;
\wop_m'=w_m'\,\R_{m,\ff}^{}\zop_m^{}\R_{m,\ff}^{-1}\;,
\end{equation}
and
\begin{equation}
\ds\uop_\ff'=u_{\ff,m+1}^{}\,
\R_{m,\ff}^{}\xop_\ff^{}\R_{m,\ff}^{-1}\;,\;\;\;
\wop_\ff'=w_{\ff,m+1}\,\R_{m,\ff}^{}\zop_\ff^{}\R_{m,\ff}^{-1}\;,
\end{equation}
Solution of (\ref{fundamental}) with respect to the primed
\emph{operators} is unique and given by
\begin{equation}\label{the-mapping}
\ds\left\{\begin{array}{l}
\ds\uop_m'\;=\;{\kappa_\ff\over\kappa}\,\uop_\ff^{}\;,\\
\\
\ds\wop_m'\;=\;
\wop_m^{}\wop_\ff^{}\,-\,\omeg\lambda_\ff^{}\uop_\ff^{-1}
\wop_\ff^{}\;,\\
\\
\ds\uop_\ff'\;=\; \left(\lambda_\ff\,+\,\kappa\uop_m\wop_m\,-\,
\omeg\wop_m\uop_\ff\right)^{-1}\lambda_\ff^{}\uop_m^{}\;,\\
\\
\ds\wop_\ff'\;=\; {\kappa\over\kappa_\ff}\,\uop_m\wop_m\,
\left(\wop_m^{}\uop_\ff^{}\,-\,\omeg\lambda_\ff^{}\right)^{-1}
\end{array}\right.
\end{equation}
$N$th powers of all primed operators in (\ref{the-mapping}) give
\begin{equation}\label{ff-recursion}
\ds u_{\ff,m+1}^N\;=\;{\lambda_\ff^N\,u_m^N\over
\lambda_\ff^N+\kappa^N u_m^N w_m^N + w_m^N u_{\ff,m}^N}\;,\;\;\;
\ds w_{\ff,m+1}^N \;=\; {\kappa^N\over\kappa_\ff^N}\,
{u_m^Nw_m^N\over \lambda_\ff^N+w_m^N u_{\ff,m}^N}\;,
\end{equation}
and
\begin{equation}\label{uw-prime}
\ds u_m^{\prime N} \;=\;
{\kappa_\ff^N\over\kappa^N}\,u_{\ff,m}^{N}\;, \;\;\;\;
w_m^{\prime N} \;=\;
{\kappa^N\over\kappa_\ff^N}\,{u_m^{N}w_m^{N}\over
u_{\ff,m}^N}\,{w_{\ff,m}^{N}\over w_{\ff,m+1}^{N}}\;.
\end{equation}
Eqs. (\ref{ff-recursion}) and (\ref{uw-prime}) are the exact and
unique solution of (\ref{LL-darboux}). Therefore eq.
(\ref{LL-darboux}) is the consistency condition for
(\ref{op-recursion}). Next, eq. (\ref{op-recursion}) is a set of
linear equations for the matrix elements of $\R_{m,\ff}$, this
set is defined appropriately, and this provides the uniqueness of
$\R_{m,\ff}$.

The local transformation
\begin{equation}\label{darboux-mapping}
\ds u_m,w_m,u_{\ff,m},w_{\ff,m}\;\;\mapsto\;\;
u_m',w_m',u_{\ff,m+1},w_{\ff,m+1}\;,
\end{equation}
given by eqs. (\ref{ff-recursion}) and (\ref{uw-prime}), is called
usually as the Darboux transformation for the classical
relativistic Toda chain. Eqs. (\ref{ff-recursion}) and
(\ref{uw-prime}) define the mapping (\ref{darboux-mapping}) up to
$N$-th roots of unity. These roots are the additional discrete
parameters of the transformation (\ref{darboux-mapping}). Note,
matrix $\R_{m,\ff}$ is unique if all these roots are fixed.

\subsection{$\mathcal{Q}$-transformation}

Relations (\ref{op-recursion}) and (\ref{LL-darboux}) may be
iterated for the whole chain. The functional counterpart
gives
\begin{equation}\label{LT-darboux}
\ds\Ldst(u_{\ff,1}^{N},w_{\ff,1}^{N})
\Montoda(\{u_m^{N},w_m^{N}\}_{m=1}^M)\;=\;
\Montoda(\{u_m^{\prime
N},w_m^{\prime N}\}_{m=1}^M)
\Ldst(u_{\ff,M+1}^{N},w_{\ff,M+1}^{N})\;,
\end{equation}
where the mapping $u_{\ff,1},w_{\ff,1}$ $\mapsto$
$u_{\ff,M+1}$,
$w_{\ff,M+1}$ is $M$-th iteration of
(\ref{darboux-mapping}).

For the periodic chain the cyclic boundary conditions for the
recursion (\ref{ff-recursion}) are to be imposed,
\begin{equation}\label{cyclic-recursion}
\ds u_{\ff,M+1}\;=\;u_{\ff,1}\;,\;\;\;w_{\ff,M+1}\;=\;
w_{\ff,1}\;.
\end{equation}
Now suppose (\ref{ff-recursion}) and (\ref{cyclic-recursion}) are
solved, i.e. $u_{\ff,m}$, $w_{\ff,m}$ are parameterized in the
terms of $u_m,w_m$, $m=1...M$, and some extra parameters, possible
degrees of freedom of (\ref{ff-recursion}) and
(\ref{cyclic-recursion}) (e.g. $\lambda_\ff$). Then
(\ref{uw-prime}) defines in general the transformation
$\mathcal{Q}_\ff$
\begin{equation}\label{Q-mapping}
\ds \mathcal{Q}_\ff\;:\;\{u_m^{},w_m^{}\}_{m=1}^M\;
\mapsto\;\{u_m',w_m'\}_{m=1}^M
\end{equation}
The transfer matrices of two sets, $\{u_m^{},w_m^{}\}$ and
$\{u_m',w_m'\}$, have the same spectrum, because of there exists
$N^M\times N^M$ matrix $\Q_\ff$, nondegenerative in general,
\begin{equation}\label{R-monodromy}
\ds \Q_\ff\;\stackrel{def}{=}\;
\textrm{tr}_{\ff}\;\left(\R_{1,\ff}\,\R_{2,\ff}\,\dots\,
\R_{M,\ff}\right)\;,
\end{equation}
such that
\begin{equation}\label{t-Q-permutation}
\ds \tp(\lambda\,;\{u_m^{},w_m^{}\})\,\cdot\, \Q_\ff\;=\;
\Q_\ff\,\cdot\, \tp(\lambda\,;\{u_m',w_m'\})\;.
\end{equation}
Subscript $\ff$ of $\Q_\ff$ stands as the reminder for the
parameters, arising in the solution of recursion, including at
least the spectral parameter $\lambda_\ff$.

Consider now the repeated application of the transformations
$\mathcal{Q}_\ff$, eq. (\ref{Q-mapping}),
\begin{equation}\label{sequence}
\ds
\{u_m^{},w_m^{}\}\stackrel{\mathcal{Q}_{\ff_1}}{\mapsto}\{u_
m',w_m'\}
\stackrel{\mathcal{Q}_{\ff_2}}{\mapsto} ...
\stackrel{\mathcal{Q}_{\ff_n}}{\mapsto}
\{u_m^{(n)},w_m^{(n)}\}\;,
\end{equation}
such that the set of isospectral quantum transfer matrices
\begin{equation}\label{transfer-set}
\ds \tp^{(n)}(\lambda) \;=\;
\tp(\lambda,\kappa;\{u_m^{(n)},w_m^{(n)}\}_{m=1}^M)\;
\end{equation}
has arisen. Sequence (\ref{sequence}) defines the transformation
$\mathcal{K}$,
\begin{equation}\label{K-mapping}
\mathcal{K}^{(n)}\;:\; \{u_m^{}\equiv u_m^{(0)},w_m^{}\equiv
w_m^{(0)}\}_{m=1}^M\;
\mapsto\;\{u_m^{(n)},w_m^{(n)}\}_{m=1}^M\;,
\end{equation}
with the finite dimensional counterpart
\begin{equation}\label{K-operator}
\ds\mathbf{K}^{(n)} \;=\;
\Q^{(1)}_{\phi_1}\,\Q^{(2)}_{\phi_2}\,...\,\Q^{(n)}_{\phi_n}
\;,
\end{equation}
where
\begin{equation}\label{Q-local}
\ds \Q^{(n)}_{\phi_n}\;=\; \textrm{tr}_{\phi_n}\;
\left(\R^{(n)}_{1,\ff_n}\,\R^{(n)}_{2,\ff_n}\,\dots\,
\R^{(n)}_{M,\ff_n}\right)
\end{equation}
makes
\begin{equation}
\ds \tp^{(n-1)}(\lambda)\;\Q^{(n)}_{\phi_n}\;=\;
\Q^{(n)}_{\phi_n}\;\tp^{(n)}(\lambda)\;,
\end{equation}
and
\begin{equation}
\ds \tp^{(0)}(\lambda)\;\mathbf{K}^{(n)}\;=\;
\mathbf{K}^{(n)}\;\tp^{(n)}(\lambda)\;.
\end{equation}

\section{Matrix $\R$}

In this section we construct explicitly the finite dimensional
matrix $\R_{m,\ff}$, obeying (\ref{op-recursion}), in the basis
(\ref{representation}). But first we have to introduce several
notations.

\subsection{$w$-function}

Let $p$ be a point on the Fermat curve $\mathcal{F}$
\begin{equation}\label{fermat}
\ds
p\;\stackrel{def}{=}\;(x,y)\;\in\mathcal{F}
\;\;\Leftrightarrow\;\;
x^N+y^N=1\;.
\end{equation}
Very useful function on the Fermat curve is $w_p(n)$,
$p\in\mathcal{F}$, $n\in\mathbb{Z}_N$, defined as follows:
\begin{equation}\label{woshka}
\ds {w_p(n)\over w_p(n-1)}\;=\;{y\over
1\,-\,x\,\omega^n}\;,\;\;\;w_p(0)\;=\;1\;.
\end{equation}

Function $w_p(n)$ has a lot of remarkable properties, see the
appendix of ref. \cite{mss-vertex} for an introduction into
$\omega$-hypergeometry. In this paper it is necessary to mention
just a couple of properties of $w$-function. Let $O$ be the
following automorphism of the Fermat curve:
\begin{equation}\label{O}
\ds p\;=\;(x,y) \;\; \Leftrightarrow \;\;
Op\;=\;(\omega^{-1}x^{-1},\omega^{-1/2}x^{-1}y)\;.
\end{equation}
Then
\begin{equation}
\ds w_{Op}(n)\;=\;{1\over \Phi(n) w_p(-n)}\;,
\end{equation}
where
\begin{equation}\label{Phi}
\Phi(n)\;=\;(-)^n\,\omega^{n^2/2}\;.
\end{equation}
In the subsequent sections we will use implicitly two simple
automorphisms else:
\begin{equation}\label{w-simp}
\ds w_{(x,\omega y)}(n)=\omega^n w_{(x,y)}(n)\;,\;\;\;
w_{(x,y)}(n+1)={y\over 1-\omega x} w_{(\omega x,y)}(n)\;.
\end{equation}
Define also three special points on the Fermat curve:
\begin{equation}\label{singular-p}
\ds q_0=(0,1)\;,\;\; q_\infty = Oq_0\;,\;\; q_1 =
(\omega^{-1},0)\;.
\end{equation}
Then
\begin{equation}\label{singular-w}
\ds w_{q_0}(n)=1\;,\;\; w_{q_\infty}(n) = {1\over\Phi(n)}
\;,\;\;
{1\over w_{q_1}(n)}\;=\;\delta_{n,0}\;.
\end{equation}

\subsection{Matrix elements of $\R_{m,\ff}$}

Consider the $N^2\times N^2$ matrix \\ $\R_{m,\ff}(p_1,p_2,p_3)$
with the following matrix elements:
\begin{equation}\label{R}
\ds\begin{array}{l}\ds
\langle\alpha_m,\alpha_\ff|\R_{m,\ff}|\beta_m,\beta_\ff\rangle\;=\\
\\
\ds =\; \omega^{(\alpha_m-\beta_m)\beta_\ff}\,
{w_{p_1}(\alpha_\ff-\alpha_m)\,w_{p_2}(\beta_\ff-\beta_m)\over
w_{p_3}(\beta_\ff-\alpha_m)}\;\delta_{\alpha_\ff,\beta_m}\;.
\end{array}
\end{equation}
Here $p_1,p_2,p_3$ are three points on the Fermat curve,
such that
\begin{equation}\label{xxxx}
\ds x_1\,x_2\;=\;x_3\;.
\end{equation}
Eq. (\ref{xxxx}) and the spin structure of (\ref{R}) provide the
dependence of (\ref{R}) on two continuous parameters, say $x_1$
and $x_3$, and on two discrete parameters, say the phase of $y_1$
and the phase of $y_3$.

\begin{prop}
Matrix $\R_{m,\ff}(p_1,p_2,p_3)$, whose matrix elements (\ref{R})
are given in the basis (\ref{representation}), makes the
following mapping:
\begin{equation}\label{finite-mapping}
\ds\left\{
\begin{array}{l}
\ds\R_{m,\ff}^{}\xop_m^{}\R_{m,\ff}^{-1}\;=\; \xop_\ff\;,\\
\\
\ds\R_{m,\ff}^{}\zop_m^{}\R_{m,\ff}^{-1}\;=\; {y_3\over
y_2}\zop_m^{}\zop_\ff^{}-\omega
{x_3y_1\over x_1y_2}\,\xop_\ff^{-1}\zop_\ff^{}\;,\\
\\
\ds\R_{m,\ff}^{}\xop_\ff^{-1}\R_{m,\ff}^{-1}\;=\; \omega
x_3\xop_m^{-1}-\omega{x_1y_3\over
y_1}\xop_m^{-1}\zop_m^{}\xop_\ff^{} +{y_3\over
y_1}\zop_m^{}\;,\\
\\
\ds \R_{m,\ff}^{}\zop_\ff^{-1}\R_{m,\ff}^{-1}\;=\; {y_3\over
y_2}\xop_\ff^{}\xop_m^{-1}- \omega {x_3y_1\over
x_1y_2}\zop_m^{-1}\xop_m^{-1}\;.
\end{array}\right.
\end{equation}
\end{prop}
Proof: direct verification with the help of (\ref{w-simp}).

Compare (\ref{finite-mapping}) with (\ref{the-mapping}).
Obviously, $\R_{m,\ff}$ solves (\ref{op-recursion}) if
\begin{equation}\label{p-uw-1}
\ds x_{1,m}\;=\;\omegg\,{u_{\ff,m}\over\kappa\,u_m}
\;,\;\;\;{x_{3,m}\,y_{1,m}\over x_{1,m}\,y_{3,m}}\;=\;
\omegg\,{\lambda_\ff^{}\over u_{\ff,m} w_m}\;,
\end{equation}
and
\begin{equation}\label{p-uw-2}
\ds\begin{array}{ll} \ds
u_m'\;=\;{\kappa_\ff\over\kappa}\,u_{\ff,m}^{}\;,& \ds
w_m'\;=\;w_m^{}w_{\ff,m}^{}\,{y_{2,m}\over y_{3,m}}\;,\\
\\
\ds u_{\ff,m+1}^{}\;=\;\omega\,x_{3,m}\,u_m^{}\;, & \ds
w_{\ff,m+1}^{}\;=\;{\kappa\over\kappa_\ff}\,{u_m\over
u_{\ff,m}}\,{y_{3,m}\over y_{2,m}}\;.
\end{array}
\end{equation}
Eqs. (\ref{p-uw-1}) and (\ref{p-uw-2}) allow one to parameterize
$\R^{(n)}_{m,\ff_n}$ in (\ref{Q-local}) in terms of
$u_m^{},w_m^{},...$. So for any appropriate set of
$\{u_m^{(k)},w_m^{(k)}\}$, $m=1...M$ and $k=0...n$, see formula
(\ref{sequence}), we may construct the corresponding
$\mathbf{K}^{(n)}$ explicitly via the parameterization
(\ref{p-uw-1},\ref{p-uw-2}), form of the matrix elements
(\ref{R}), and the formulas (\ref{Q-local},\ref{K-operator}).

Further we will consider the sequences like (\ref{sequence}) for
the homogeneous initial state $\{u_m=u_0,w_m=w_0\}_{m=1}^M$, and
parameterize $\{u_m^{(n)},w_m^{(n)}\}$ in the terms of solitonic
Hirota-type expressions. To do it, several notations are to be
introduced.

\section{Rational $\Theta$-function}

In this section we introduce several useful functions and
notations. The main purpose is to introduce the form of solitonic
$\tau$-function for the classical relativistic Toda chain. The
reader may find the details concerning the corresponding
classical integrable model, reduction from 2DToda lattice
hierarchy and so on in Ref. \cite{sp-3}.

Fix any sequence of $\phi_k$ from the set $\Set$:
\begin{equation}\label{phi-k}
\ds\phi_k\;\in\;\Set\;\stackrel{def}{=}\; \{{\,\pi\over
M}\,,\;{2\pi\over M}\,,\;...\;,\;{(M-1)\pi\over M}\,\}\;,
\end{equation}
Introduce the functions
\begin{equation}\label{phi-parametrization}
\ds\left\{\begin{array}{l} \ds
\Om_\phi^{}\;=\;\EXP^{i\,\phi}\,\left(
\sqrt{\cos^2\phi\,+\,\kappa^N}\;+\;\cos\phi\right)\;,\\
\\
\Om^*_\phi\;=\;\EXP^{-i\,\phi}\,\left(
\sqrt{\cos^2\phi\,+\,\kappa^N}\;+\;\cos\phi\right)\;,\\
\\
\ds \Lambda_\phi\;=\;\Om^{}_\phi\;\Om^*_\phi\;.
\end{array}\right.
\end{equation}
Expressions for $\Delta$ and $\Delta^*$ parameterize the curve
\begin{equation}\label{Delta-curve}
\ds\Delta\Delta^*\;=\;\Delta\,+\,\Delta^*\,+\,\kappa^N\;\;
\textrm{in the terms of}\;\; \ds
\EXP^{2i\phi}\;=\;{\Delta\over\Delta^*}\;.
\end{equation}

Rational $\Theta$-functions are to be defined recursively as
\begin{equation}
\ds\begin{array}{l} \ds \Theta_m^{(0)}\;=\;1\;,\\
\\
\ds \Theta_m^{(1)}\;=\;1\,-\,f_1\,\EXP^{2 i m \phi_1}\;,\\
\\
\ds \Theta_m^{(2)}\;=\; 1-f_1\EXP^{2 i m \phi_1}- f_2\EXP^{2
i m
\phi_2} + d_{1,2} f_1 f_2 \EXP^{2 i m (\phi_1+\phi_2)}\;,
\end{array}\end{equation}
and so on,
\begin{equation}\label{Tau}
\ds
\Theta_m^{(n)}\;=\;\Theta_m^{(n-1)}(\{f_k,\phi_k\}_{k=1}^{n-
1})-
f_n\EXP^{2 i m
\phi_n}\Theta_m^{(n-1)}(\{d_{k,n}f_k,\phi_k\}_{k=1}^{n-1})\;
,
\end{equation}
where all $\phi_k$ are different and belong to the set
(\ref{phi-k}). The phase shift $d_{j,k}=d_{\phi_j,\phi_k}$ is
given by (see (\ref{phi-parametrization}))
\begin{equation}\label{phase-shift}
\ds d_{\phi,\phi'}\;=\;
{\left(\Om^{}_{\phi}-\Om^{}_{\phi'}\right)\;
\left(\Om^*_{\phi}-\Om^*_{\phi'}\right) \over
\left(\Om^{}_{\phi}-\Om^*_{\phi'}\right)\;
\left(\Om^*_{\phi}-\Om^{}_{\phi'}\right) }\;.
\end{equation}
Arguments of $\Theta_m^{(n)}$ are the set of $\phi_k\in\Set$ and
corresponding amplitudes $f_k$, $k=1...n$. Maximal
$\Theta$-function is
\begin{equation}
\ds \Theta_m^{}\;\stackrel{def}{=}\;
\Theta_m^{(M-1)}(\{f_k\}_{k=1}^{M-1})\;.
\end{equation}
$\Set$-set, (\ref{phi-k}), provides the $M$-periodicity of
$\Theta$-functions, $\ds\Theta_{m+M}^{}\equiv\Theta_m^{}$.

Introduce next a couple of useful functions
\begin{equation}\label{c-function}
\ds s_\phi(\xi)\;=\; {\Om^*_{\phi}\over\Om^{}_{\phi}}\;
{\Om^{}_{\phi}\,-\,\xi\over \Om^*_{\phi}\,-\,\xi}\;,
\end{equation}
and
\begin{equation}\label{s-function}
\ds
s_{\phi,\phi'}\;=\;s_\phi(\Om^*_{\phi'})\;=\;
{\Om^*_{\phi}\over\Om^{}_{\phi}}\;{\Om^*_{\phi'}
\,-\,\Om^{}_{\phi}\over \Om^*_{\phi'}\,-\,\Om^*_{\phi}}\;.
\end{equation}

Necessary is to consider the re-parameterization of the amplitudes
$f_k$ in the special form: for any $n,k$, $0\leq n < M$, $1\leq
k\leq n$ let
\begin{equation}\label{free-amplitudes}
\ds f_k^{(n)}(\{g\})\;=\;g_k\;\prod_{j=1\atop j\neq
k}^n\;s_{k,j}\;,
\end{equation}
where $s_{k,j}$ is given by (\ref{s-function}). Denote the
corresponding $\Theta$-functions as follows:
\begin{equation}\label{free-theta}
\ds \Theta_{m}^{(n)}(\{g\})\;=\;
\Theta_m^{(n)}(\{f_k^{(n)}(\{g\}),\phi_k\}_{k=1}^n)
\end{equation}
In advance, we will always imply the free amplitudes $g_k$,
$k=1,...$, as the arguments of $\Theta$-function,
(\ref{free-amplitudes},\ref{free-theta}).

Particular case when all $g_k=s_k(\xi)$ is also important: let
\begin{equation}\label{xi-theta}
\ds \Theta_{m}^{(n)}(\xi)\;=\;
\Theta_m^{(n)}(\{g_k=s_k(\xi)\}_{k=1}^{n})
\end{equation}
This particular case has the following property:
\begin{prop}\label{prop-values}
For any $n$
\begin{equation}
\ds \Theta^{(n)}_{n}(\xi)\;=\;
\prod_{k=1}^{n}\,{\Om^*_k\,-\,\Om^{}_k\over
\Om^*_k\,-\,\xi}\;
\prod_{1\leq j < k \leq n}\; {\Lambda_j\,-\,\Lambda_k\over
\Om^*_j\,-\,\Om^*_k}\;,
\end{equation}
and for $m=0...n$
\begin{equation}\label{theta-xi-values}
\ds \Theta^{(n)}_{m}(\xi)\;=\;
\left(-\,\kappa^N\right)^{(n-m)(n-m-1)/2}\;
\left({\xi\over\Om_1...\Om_n}\right)^{n-m}\;\Theta^{(n)}_{n}
(\xi)
\end{equation}
If $\xi=1$, eq. (\ref{theta-xi-values}) is valid also for $m=-1$.
\end{prop}

\section{Parameterization of $\mathbf{K}$}

\subsection{Sequence of the B\"acklund transformations}

Turn now to the original quantum chain. Let the initial
parameters $u_m,w_m$ are homogeneous:
\begin{equation}\label{uw-homo}
\ds u_m\;=\;-\omega^{-1/2}\;,\;\;w_m=-1\;.
\end{equation}
It corresponds to (\ref{gauge-fix}).

For the sequence of $\phi_k\in\Set$ fixed, and for generic set of
corresponding $g_k$, $k=1...M-1$, let (see eqs.
(\ref{free-amplitudes}) and (\ref{free-theta}) and the note right
after (\ref{free-theta}))
\begin{equation}\label{tau-theta}
\ds
\left(\tau_m^{(n)}\right)^N\;=\;
\Theta_m^{(n)}(\{g_k\}_{k=1}^{n})\;,\;
\;\;
\left(\theta_m^{(n)}\right)^N\;=\;
\Theta_m^{(n)}(\{g_ks_k(1)\}_{k=1}^{n})\;.
\end{equation}
The phases of $\tau_m^{(n)}$ and $\theta_m^{(n)}$ are
arbitrary.

\begin{prop}\label{prop-inhomo}
Consider the set of $(M-1)$ transformations
(\ref{Q-mapping}),
such that the initial state is the homogeneous one
(\ref{uw-homo}), and the mapping $\mathcal{Q}_{\phi_n}$ is
derived for
\begin{equation}\label{lambda-spectrum}
\ds \lambda_{\phi_n}^N\;=\;\Lambda_{\phi_n}^{}\;,
\end{equation}
where sequence of $\phi_n\in\Set$ is fixed, and function
$\Lambda_\phi$ is given by the last formula of
(\ref{phi-parametrization}). Let besides
\begin{equation}\label{kappa-ff-value}
\ds \kappa_{\phi_n}^N\;=\;{\kappa^N\over\Delta_{\phi_n}}\;.
\end{equation}
Then the sequence (\ref{sequence}) of
$\{u^{(n)}_m,w^{(n)}_m\}$
may be parameterized as follows:
\begin{equation}\label{uw-inhomo}
\ds
u_m^{(n)}\;=\;-\omega^{-1/2}\,{\tau^{(n)}_{m-1}\over\tau^{(n
)}_m}\;,
\;\;\;
w_m^{(n)}\;=\;-\,{\theta^{(n)}_m\over\theta^{(n)}_{m-1}}\;,\
;\;\;
m\,\in\,\mathbb{Z}_M\;,
\end{equation}
where $\tau^{(n)}_m$, $\theta^{(n)}_m$ are given by
(\ref{tau-theta}) with arbitrary $g_1...g_{M-1}$.
\end{prop}
See \cite{sp-1} for the sketch of the proof of this proposition.
Eq. (\ref{kappa-ff-value}) needs a comment. Arbitrary value of
$\kappa_{\phi_n}$ corresponds to a gauge transformation of
\begin{equation}
\ds \{u_m^{(n)},w_m^{(n)}\}_{m=1}^M\;\mapsto\; \{ c u_m^{(n)},
c^{-1} w_m^{(n)}\}_{m=1}^M
\end{equation}
with some $c$ proportional to $\kappa_{\phi_n}$, while the
structure of $\tau$ and $\theta$ is not changed. So we impose
(\ref{kappa-ff-value}) and obtain (\ref{uw-inhomo}) without lost
of generality.

\subsection{Parameterization of modified $Q$-operators}

Let now for the shortness $\lambda_{\phi_n}=\lambda_n$ are given
by (\ref{lambda-spectrum}), and let us fix further the roots for
$\delta_n$ and $c_n$, $1\leq n < M$ :
\begin{equation}\label{delta-kappa-c}
\ds \delta_n^N=\Om_{\phi_n}^{}\;,\;\;\;
\kappa_{\phi_n}={\kappa\over\delta_n}\;,\;\;\;
 c_n^N={\Om^*_{\phi_n}-1 \over \Om^*_{\phi_n}}\;.
\end{equation}
Auxiliary values in (\ref{p-uw-1},\ref{p-uw-2}) are given by
\begin{equation}
\ds u_{\phi_n,m} \;=\;
-\,\omegg\,\delta_n\,{\tau_{m-1}^{(n)}\over\tau_m^{(n)}}\;, \;\;\;
w_{\ff_n,m+1}\;=\; \omega^{-1/2}{\delta_n\over\lambda_n c_n}\,
{\theta_m^{(n-1)}\tau_m^{(n)}\over\tau_m^{(n-1)}\theta_m^{(n)}}\;.
\end{equation}

$\R^{(n)}_{m,\ff_n}$-matrices, entering  to (\ref{Q-local}),have
the arguments
\begin{equation}
\ds \R_{m,\phi_n}^{(n)}\;:\;p_{1,m}^{(n)},\;p_{2,m}^{(n)},
\;p_{3,m}^{(n)},
\end{equation}
given by
\begin{equation}\label{xy-explicit}
\ds\begin{array}{ll} \ds x_{1,m}^{(n)}\;=\;
\omega^{-1/2}\,{\delta_n\over\kappa}\,
{\tau_{m\phantom{1}}^{(n-1)}\tau_{m-1}^{(n)}
\over\tau_{m-1}^{(n-1)}\tau_{m\phantom{1}}^{(n)}}\;,& \ds
x_{3,m}^{(n)}\;=\;\omega^{-1}\delta_n^{}\,
{\tau_{m\phantom{1}}^{(n-1)}\tau_{m\phantom{1}}^{(n)}\over
\tau_{m-1}^{(n-1)}\tau_{m+1}^{(n)}}\;,\\
&\\
\ds {y_{3,m}^{(n)}\over
y_{2,m}^{(n)}}\;=\;\omega^{-1/2}\,{\delta_n\over\lambda_n c_n}\,
{\theta_{m\phantom{1}}^{(n-1)}\tau_{m-1}^{(n)}\over
\tau_{m-1}^{(n-1)}\theta_{m\phantom{1}}^{(n)}}\;, & \ds
{y_{1,m}^{(n)}\over y_{3,m}^{(n)}}\;=\;
\omega^{1/2}\,{\lambda_n\over\kappa\delta_n}\,
{\theta_{m-1}^{(n-1)}\tau_{m+1}^{(n)}\over
\theta_{m\phantom{1}}^{(n-1)}\tau_{m\phantom{1}}^{(n)}}\;,
\end{array}\end{equation}
and with (\ref{xxxx}) implied,
\begin{equation}
\ds x_{2,m}^{(n)}\;=\;
\omega^{-1/2}\kappa\,{(\tau_m^{(n)})^2\over\tau_{m-1}^{(n)}
\tau_{m+1}^{(n)}}\;.
\end{equation}

The trace of quantum monodromy (\ref{Q-local}) may be calculated
explicitly in the basis (\ref{representation}), and the answer is
\begin{equation}\label{Q-elements}
\ds \langle\alpha|\mathbf{Q}^{(n)}_{\phi_n}|\beta\rangle =
\prod_{m\in\mathbb{Z}_M}\;
\omega^{(\alpha_m-\beta_m)\beta_{m+1}}\,
{w_{p_{1,m}^{(n)}}(\beta_m-\alpha_m)\,
w_{p_{2,m}^{(n)}}(\beta_{m+1}-\beta_m)\over
w_{p_{3,m}^{(n)}}(\beta_{m+1}-\alpha_m)}\;.
\end{equation}
Operator $\mathbf{K}$, (\ref{K-operator}), calculated as the
product of all $(M-1)$ $Q^{(n)}$-operators, is thus defined
explicitly.
\begin{prop}
Operator $\mathbf{K}(\{g\})$, given by (\ref{K-operator}), does
not depend on the ordering of $(g_k,\phi_k)$ (up to a multiplier,
arising in general when one takes arbitrary phases for
$\tau_{m}^{(n)}$ and $\theta_m^{(n)}$). Thus without lost of
generality one may regard $\ds \phi_n\;=\;{\pi n\over M}$.
\end{prop}
Thus for the generic set of amplitudes $g_k$ constructed is
the
similarity operator $\mathbf{K}(\{g\})$, transforming the
homogeneous initial transfer matrix into most generic
isospectral
one.

\subsection{Arbitrary value of $\lambda_\ff$}

Previously  we considered the set of $\phi_k\in\Set$. Actually,
all the calculations in the classical relativistic Toda chain are
based on (\ref{Delta-curve}). Therefore, in general, we may
consider a generic sequence of complex numbers $\phi_k$,
$k\in\mathbb{Z}_+$. For any generic complex $\phi$ one has to
restore $\lambda_\phi$ and $\delta_\phi$ via
(\ref{phi-parametrization}), and take into account
(\ref{delta-kappa-c}). It corresponds to the $Q$ operator for the
generic value of $\lambda_\ff$. Then all the formulas for the
parameterization of transfer matrix in the terms of
$\tau$-functions
(\ref{uw-inhomo},\ref{tau-theta},\ref{free-theta}) as well as the
parameterization of $Q$-operator
(\ref{xy-explicit},\ref{Q-elements}) are valid if the amplitudes
$g_\phi$, corresponding to those $\phi\not\in\Set$ and
$-\phi\not\in\Set$ are zeros, and in the case when $\phi\in\Set$
or $-\phi\in\Set$ the amplitudes are arbitrary. Important is that
in the calculation of the amplitudes (\ref{free-amplitudes}) all
$\phi$s are to be taken into account. Also, if in the generic
sequence of $\phi_k$ it happens $\phi_k=\phi_m$ or
$\phi_k=-\phi_m$, then $\Theta$-function is to be understood as
the corresponding residue of formal singular expression
(\ref{free-theta}).

For example, usual $Q$-operators for the homogeneous initial state
corresponds to $\phi_1=\phi\not\in\Set$, $g_1=0$. The cases when
$\phi_k=\pm\phi_m$ correspond to the annihilation of the solitons.

Nevertheless, considering operator $\mathbf{K}$,
(\ref{K-operator}), we will always imply $\phi_k\in\Set$.

\subsection{Baxter equation}

Consider initial homogeneous transfer matrix $\tp(\lambda)$ and
corresponding $Q$ operators with arbitrary value of $\lambda_\ff$.
As it was mentioned in the previous subsection, corresponding
functional counterpart of $Q$-operator is trivial, i.e. such
$Q$-operators commute with the transfer matrices and form the
commutative family. Such $Q$-operators are usual Baxter's
$Q$-operators, obeying Baxter's equation.

Explicit form of simple $Q$-operator is given by
(\ref{Q-elements}) with the homogeneous parameterization
\begin{equation}
\ds p_{1,m}=p_1\;,\;\;p_{2,m}=p_2\;,\;\;p_{3,m}=p_3
\end{equation}
with
\begin{equation}\label{xy-homo}
\ds\begin{array}{l} \ds
x_1=\omega^{-1/2}{\delta_\ff\over\kappa}\;,\;\;
x_2=\omega^{-1/2}\kappa\;,\;\;x_3=\omega^{-1}\delta_\ff\;,\\
\\
\ds {y_3\over y_2}={\omega^{-1/2}\delta_\ff\over\lambda_\ff
c_\ff}\;,\;\; {y_1\over
y_3}={\omega^{1/2}\lambda_\ff\over\kappa\delta_\ff}\;,
\end{array}
\end{equation}
where $\delta_\ff$, $\lambda_\ff$ and $c_\ff$ are given by
(\ref{delta-kappa-c},\ref{lambda-spectrum}) and
(\ref{phi-parametrization}), but with generic value of $\phi$.
Using (\ref{xy-homo}), we may regard
\begin{equation}
\ds \mathbf{Q}\;=\;\mathbf{Q}(\lambda_\ff,\delta_\ff)\;.
\end{equation}
Several simple properties of simple $Q$-operator may be derived
with the help of the matrix elements of (\ref{Q-elements}) and
(\ref{xy-homo}), and with the help of (\ref{w-simp}):
\begin{equation}\label{Q-prop}
\ds
\begin{array}{l}
\ds
\mathbf{X}\mathbf{Q}(\lambda_\ff,\delta_\ff)\mathbf{X}^{-1}\;=\;
\mathbf{Q}(\omega^{-1}\lambda_\ff,\delta_\ff)\;,\\
\\
\ds \mathbf{Z}\mathbf{Q}(\lambda_\ff,\delta_\ff)\mathbf{X}\;=\;
\left(\omega^{1/2}{\lambda_\ff\over\delta_\ff}\,{1-\delta_\ff\over
\kappa-\omega^{1/2}\delta_\ff}\right)^M\,\mathbf{Q}(\omega\lambda_\ff,\omega\delta_\ff)\;,\\
\\
\ds
\mathbf{Y}\mathbf{Q}(\lambda_\ff,\delta_{\ff})\;=\;\mathbf{Q}(\lambda_\ff,\delta_\ff)\mathbf{Y}\;=\;
\left(-{\lambda_\ff\over\delta_\ff}\,{1-\delta_\ff\over\kappa-\omega^{1/2}\delta_\ff}\right)^M\,
\mathbf{Q}(\lambda_\ff,\omega\delta_\ff)\;,
\end{array}
\end{equation}
where
\begin{equation}\label{XZ}
\ds\mathbf{X}=\prod_{m=1}^M\,\xop_m\;,\;\;\;
\mathbf{Z}=\prod_{m=1}^M\zop_m\;,
\end{equation}
and $\mathbf{Y}$ is given by (\ref{Y}).

Let us derive now Baxter's equation for general inhomogeneous
chain. To do it, one has to consider the degeneration point of
$\ldst$, (\ref{ldst-uwkappa}):
\begin{equation}\label{degen}
\ds\ldst_\ff(\lambda_\ff,\lambda_\ff)\;=\;
{1-\omega^{1/2}\kappa_\ff^{}\wop_\ff^{} \choose
-\omega^{1/2}\lambda_\ff^{}\uop_\ff^{-1}\wop_\ff^{}}\,\cdot\,
\biggl(\;1\;,\;-\omega^{1/2}\lambda_\ff^{-1}\uop_\ff^{}\biggr)\;.
\end{equation}
Therefore eq. (\ref{op-recursion}) in the degeneration point may
be rewritten in the following forms:
\begin{equation}\label{degen-1}
\ds\begin{array}{l} \ds
\biggl(\;1\;,\;-\omeg{u_{\ff,m}\over\lambda_\ff}\,\xop_\ff^{
}\;\biggr) \,\cdot\,
\ltoda_m(\lambda_\ff\,;u_m^{},w_m^{})\,\cdot\,
\R_{m,\ff}^{}\;=\\
\\
\ds=\; \R_{m,\ff}^{(1)}\,\cdot\,
\biggl(\;1\;,\;-\omeg{u_{\ff,m+1}\over\lambda_\ff}\,\xop_\ff
^{}\;\biggr)\;,
\end{array}\end{equation}
and
\begin{equation}\label{degen-2}
\ds\begin{array}{l} \ds
\ltoda_m(\lambda_\ff\,;u_m',w_m')\,\cdot\,\R_{m,\ff}
\,\cdot\,\left(\begin{array}{c}
\ds\omeg{u_{\ff,m+1}\over\lambda_\ff}\xop_\ff^{}\\
\\
\ds 1 \end{array}\right) \;=\\
\\
\ds =\; \left(\begin{array}{c}
\ds\omeg{u_{\ff,m}\over\lambda_\ff}\xop_\ff^{}\\
\\
\ds  1
\end{array}\right)\,\cdot\,\R_{m,\ff}^{(2)}\;,
\end{array}\end{equation}
where
\begin{equation}
\ds\begin{array}{l} \ds \R_{m,\ff}^{(1)}\;=\; {u_m\over
u_{\ff,m+1}}\;\;\xop_m^{}\;\R_{m,\ff}\;\xop_\ff^{-1}\;,\\
\\
\ds \R_{m,\ff}^{(2)}\;=\;
\omeg\,{u_{\ff,m+1}\,w_{m}\over\lambda_\ff}\;\;
\zop_m^{}\;\R_{m,\ff}\;\xop_\ff^{}\;.
\end{array}\end{equation}
Let further $\Q^{(1)}_\ff$ and $\Q^{(2)}_\ff$ are the traces of
the monodromies of $\R_{m,\ff}^{(1)}$ and $\R_{m,\ff}^{(2)}$
($\mathbb{Z}_M$ boundary condition for $u_{\ff,m}$ is taken into
account again -- the subscript $\ff$ reminds this). Then from
(\ref{degen-1}) and (\ref{degen-2}) it follows the Baxter equation
in its operator form:
\begin{equation}\label{BE-operator-12}
\ds \tp(\lambda_\ff)\;\Q_\ff\;=\; \Q_\ff\;\tp'(\lambda_\ff)\;=\;
\Q^{(1)}_\ff\;+\;\Q^{(2)}_\ff\;.
\end{equation}
Using the first relation of (\ref{finite-mapping}), one may obtain
\begin{equation}
\ds \Q^{(1)}_\ff=\prod_m{u_m\over
u_{\ff,m}}\;\mathbf{X}\Q\mathbf{X}^{-1}\;,\;\;\;
\Q^{(2)}=\prod_{m}\omega^{1/2}{u_{\ff,m}w_m\over\lambda_\ff}\;
\mathbf{Z}\mathbf{Q}\mathbf{X}\;,
\end{equation}
where $\mathbf{X}$ and $\mathbf{Z}$ are given by (\ref{XZ}).

Turn now to the homogeneous chain. Using (\ref{Q-prop}), we obtain
\begin{equation}\label{BE-operator}
\ds\tp(\lambda_\ff)\Q(\lambda_\ff)=
\delta_\ff^{-M}\Q(\omega^{-1}\lambda_\ff)\,+\,
\left(-\omega^{-1/2}{\delta_\ff\over\lambda_\ff}\right)^M\mathbf{Y}\Q(\omega\lambda_\ff)\;,
\end{equation}
where $\delta_\ff$-argument of all $\Q$-s remains unchanged, and
$\tp$ and $\Q$ may be diagonalized simultaneously. Let
$t(\lambda)$ and $q_t(\lambda)$ be the eigenvalues of $\tp$ and
$\Q$ for the same eigenvector, then (\ref{BE-operator}) provides
the functional equation
\begin{equation}\label{BE}
\ds t(\lambda) q_t(\lambda)= \delta^{-M}
q_t(\omega^{-1}\lambda)\,+\,
\left(-\omega^{-1/2}{\delta\over\lambda}\right)^M\omega^\gamma
q_t(\omega\lambda)\;,
\end{equation}
where $\omega^\gamma$ is the eigenvalue of $\mathbf{Y}$, and $q_t$
is the meromorphic function on the curve (\ref{curve}). Detailed
investigations of (\ref{BE}) show that for generic $\lambda$ any
solution of (\ref{BE}) such that $t(\lambda)$ is a polynomial and
$q_t(\lambda,\delta)$ is a meromorphic function on the curve
(\ref{curve}), gives the eigenvalue of $\tp(\lambda)$ and
$\mathbf{Q}(\lambda)$.

\section{Separating operator}

\subsection{Spectrum of  $\mathbf{b}$}

For the homogeneous chain (\ref{uw-homo}) the value of
$B(\lambda^N)$, (\ref{Mon-toda}), may be calculated
immediately:
\begin{equation}\label{B-spectrum}
\ds B(\lambda^N)\;=\;(-)^{N-1}\,{1\over\lambda^N}\,
\prod_{k=1}^{M-1}\,
\left(1\,-\,{\Lambda_{\phi_k}\over\lambda^N}\right)\;,
\end{equation}
where $\Lambda_{\phi}$ is given by the last equation of
(\ref{phi-parametrization}) and $\phi_k$ is exactly the set
$\Set$, (\ref{phi-k}). This defines the parameterization of the
vectors $|\{\lambda_k\}_{k=1}^{M-1},\gamma\rangle$, see
(\ref{b-eigen}).

In \cite{sp-1} it was established the following fact:
\begin{prop}
Let $\lambda_k=\lambda_{\phi_k}$ are defined by
(\ref{lambda-spectrum}) and  the phases of $\lambda_k$ are fixed.
Operator $\mathbf{K}_{\{\lambda\}}$, corresponding to the case
when all $g_k=1$, $k=1...M-1$ in eqs.
(\ref{tau-theta},\ref{free-theta}), makes the quantum separation
of variables:
\begin{equation}
\ds \mathbf{b}(\lambda_k)\,\mathbf{K}_{\{\lambda\}}\;=\;0\;,
\;\;\; k=1...M-1\;.
\end{equation}
\end{prop}
This means
\begin{equation}\label{K-separation}
\ds \mathbf{K}_{\{\lambda\}}\;=\;\sum_\gamma\;
|\{\lambda_k\}_{k=1}^{M-1},\gamma\rangle\,\langle\chi_\gamma
|\;,
\end{equation}
where $\chi_\gamma$ are some vectors, such that
$\langle\chi_\gamma|\,\mathbf{Y}=\omega^\gamma\,\langle\chi_
\gamma|$, where $\mathbf{Y}$ is given by (\ref{Y}).
Parameterization
\begin{equation}\label{sep}
\ds \left(\tau_m^{(n)}\right)^N\;=\;
\Theta_m^{(n)}(0)\;,\;\;\;
\left(\theta_m^{(n)}\right)^N\;=\;\Theta_m^{(n)}(1)
\end{equation}
is not appropriate for (\ref{xy-explicit}), because of
$\tau^{(n)}_m=0$ for $m=0...n-1$, and $u_m^{(n)}$ are ambiguous.
Thus one needs to define in addition a limiting procedure for
(\ref{sep}). In general, the ambiguity in (\ref{uw-inhomo})
corresponds to the arbitrariness of the right eigenstates
$\chi_\gamma$ of $\mathbf{K}_{\{\lambda\}}$.

\subsection{Limiting procedure}

In this section we describe an appropriate limiting procedure.
Most simple $\mathbf{K}_{\{\lambda\}}$ appears when we choose
$u^{(M-1)}_m=0$ for $m=1...M-1$. Namely, consider the set of
infinitely small numbers,
\begin{equation}
\ds\varepsilon_k\,\mapsto\,0\;,\;\;\; k=1...M-1\;,
\end{equation}
such that any ratio $\varepsilon_n/\varepsilon_m\neq 1$ is
finite, and
\begin{equation}
\ds u^{(M-1)}_m\;=\;-\omega^{-1/2}\, {\varepsilon_{M-m}\over
\delta_1....\delta_{M-1}}\;,\;\;\;m=1...M-1\;,
\end{equation}
so that
\begin{equation}
\ds u^{(M-1)}_M\;=\;-\omega^{-1/2}\,
{(\delta_1...\delta_{M-1})^{M-1} \over
\varepsilon_1...\varepsilon_{M-1}}\;.
\end{equation}
Recursion (\ref{ff-recursion},\ref{uw-prime}) with
\begin{equation}
\ds\kappa_{\phi_n}\;=\;{\kappa\over\delta_n}
\end{equation}
implies
\begin{equation}
\ds u_m^{(n)}\;=\;-\omega^{-1/2}\, {\varepsilon_{n-m+1}\over
\delta_1...\delta_n}\,+\,o(\varepsilon_1)\;,\;\;\; m=1...n
\end{equation}
while for $m=n+1...M-1$ all $u_m^{(n)}$ are regular. This
means
\begin{equation}
\ds\tau^{(n)}_m\;=\;\tau^{(n)}_n\;
{\varepsilon_1...\varepsilon_{n-m} \over
(\delta_1...\delta_n)^{n-m}}\,+\,o(\varepsilon_1^{n-m})\;,\;
\;\;
m=0...n\;.
\end{equation}

$w_m^{(n)}$ as explicit functions of $\varepsilon_k$ are rather
complicated, and speaking truly, we may say nothing about
$\tp^{(M-1)}$ in the meantime. But in the limit $\varepsilon_k =
0$ the following formulas may be chosen:
\begin{equation}
\ds {\tau^{(n)}_n\over \theta^{(n)}_n}\;=\;
\prod_{k=1}^n\;c_k\;,
\end{equation}
where $c_k$ are given by (\ref{delta-kappa-c}) and
\begin{equation}\label{simple-theta}
\ds \theta^{(n)}_m\;=\;
(\omega^{1/2}\kappa)^{(n-m)(n-m-1)/2}
{\theta^{(n)}_n\over (\delta_1...\delta_n)^{n-m}}\;.
\end{equation}
On $N$-th powers these formulas follow from proposition
(\ref{prop-values}). Eq. (\ref{simple-theta}) implies in
part
\begin{equation}\label{w/w}
\ds {w_{m+1}^{(n)}\over
w_{m}^{(n)}}\;=\;\omega^{1/2}\kappa\;,\;\;\; m=0...n-1\;.
\end{equation}

Now in the limit $\varepsilon_k\mapsto 0$ the following
parameterization of the ambiguous and singular points arises
(notation for the singular points are described by
(\ref{singular-p},\ref{singular-w})):
\begin{itemize}
\item \underline{$m=1...n-1$}:
\begin{equation}
\ds p^{(n)}_{3,m}\;=\;q_1\;,\;\;\;
p^{(n)}_{2,m}\;=\;O(p^{(n)}_{1,m})\;.
\end{equation}
where $\ds x_{1,m}^{(n)}\;=\; {1\over
\omega^{1/2}\kappa}\,{\varepsilon_{n-m+1}\over\varepsilon_{n
-m}}$.
\item \underline{$m=n$}:
\begin{equation}
\ds p_{1,n}^{(n)}=q_0\;,\;\;\;
p_{2,n}^{(n)}=q_\infty\;\;\;\forall\;n
\end{equation}
and $p_{3,n}^{(n)}$ are regular, except
\begin{equation}
\ds p_{3,M-1}^{(M-1)}=q_\infty\;.
\end{equation}
\item \underline{$m=n+1...M-2$}:
All $p_{j,m}^{(n)}$ in this region are regular.
\item \underline{$m=M-1$}:
$p_{1,M-1}^{(n)}$ are regular except
\begin{equation}
\ds p_{1,M-1}^{(M-1)}=q_0\;,
\end{equation}
and
\begin{equation}
\ds
p_{2,M-1}^{(n)}=p_{3,M-1}^{(n)}=q_\infty\;\;\;\forall\;n\;.
\end{equation}
\item \underline{$m=M$}:
\begin{equation}
\ds p_{1,M}^{(n)}=q_\infty\;,\;\;\; p_{2,M}^{(n)} =
p_{3,M}^{(n)}=q_0\;\;\;\forall\;n.
\end{equation}
\end{itemize}
Substituting these expressions into (\ref{Q-elements}), one
obtains the following form of $n$-th modified $Q$-matrix, $n\neq
M-1$:
\begin{equation}\label{Q-n}
\ds\begin{array}{l} \ds
\langle\alpha|\mathbf{Q}^{(n)}|\beta\rangle\;=\;
{1\over\Phi(\beta_1+\beta_M-\alpha_M)}\,\prod_{m=1}^{n-1}\,
\delta_{\alpha_m,\beta_{m+1}}\times\\
\\
\ds
\Phi(\alpha_n)\,w_{Op^{(n)}_{3,n}}(\alpha_n-\beta_{n+1})\times\\
\\
\ds \prod_{m=n+1}^{M-2}\,\omega^{(\alpha_m-\beta_m)\beta_{m+1}}\,
{w_{p_{1,m}^{(n)}}(\beta_m-\alpha_m)\,
w_{p_{2,m}^{(n)}}(\beta_{m+1}-\beta_m)\over
w_{p_{3,m}^{(n)}}(\beta_{m+1}-\alpha_m)}\times\\
\\
\ds
{\Phi(\alpha_{M-1})\over\Phi(\beta_{M-1})}\,
w_{p_{1,M-1}^{(n)}}(\beta_{M-1}-\alpha_{M-1})\;,
\end{array}
\end{equation}
and for the last $(M-1)$th one
\begin{equation}\label{Q-last}
\ds\langle\alpha|\mathbf{Q}^{(M-1)}|\beta\rangle\;=\;
{1\over\Phi(\beta_1+\beta_M-\alpha_M)}\,\prod_{m=1}^{M-2}\,
\delta_{\alpha_m,\beta_{m+1}}\,
\Phi(\alpha_{M-1})\;.
\end{equation}
Here $\Phi$ is given by (\ref{Phi}). Explicit form of the
modified $Q$-operators (\ref{Q-n}) and (\ref{Q-last}) allows one
to prove the following
\begin{prop}
\begin{equation}
\ds
\mathbf{K}_{\{\lambda\}}\,(\zop_{m}-\zop_{m+1})\;=\;0\;,\;\;
\;m\,\in\,\mathbb{Z}_M\;.
\end{equation}
\end{prop}

So, in our limiting procedure $\chi$ is simple:
\begin{equation}
\ds \langle\chi_\gamma|\alpha\rangle\;=\;
\chi_\gamma(\overline{\alpha})
\;\equiv\;\chi_\gamma(\overline{\alpha}\mod N),
\end{equation}
where
\begin{equation}
\ds\overline{\alpha}\;\stackrel{def}{=}
\;\sum_{m\in\,\mathbb{Z}_M}\,\alpha_m
\end{equation}
and
\begin{equation}
\ds
{\chi_\gamma(\overline{\alpha}+M)\over
\chi_\gamma(\overline{\alpha})}\,=\,
(-\omega^{-1/2})^M\,\omega^{\gamma-\overline{\alpha}}\;.
\end{equation}

\section{Discussion}

In this paper we investigated the simplest integrable model,
associated with the local Weyl algebra at the root of unity. All
such models always contain the classical discrete dynamic of
parameters. Nontrivial solution of the classical counterpart
provides the solution of the isospectrality problem of the finite
dimensional counterpart. Well known result by  Sklyanin, Kuznetsov
at al was that in the classical limit of the usual Toda chain (and
many other models) Baxter's quantum $Q$-operator corresponds to
the B\"acklund transformation of the classical system, see e.g.
\cite{Sklyanin-sep,Sklyanin-rev,KS-manybody,Sklyanin-fun,kss-dst}
etc. In our case we have the B\"acklund transformation of the
classical counterpart and modified $Q$-operator in the quantum
space simultaneously. Unusual is that solving the quantum
isospectrality problem, we miss the commutativity of the modified
$Q$-operators.

Nevertheless, our main result, we think, is the explicit
construction of $(M-1)$-parametric family of quantum inhomogeneous
transfer matrices with the same spectrum as the initial
homogeneous one, and the explicit construction of the
corresponding similarity operator (\ref{K-operator}). We hope, the
solution of the isospectrality problem will help to solve the
model with arbitrary $N$ explicitly.

As one particular application of it we have obtained the quantum
separation of variables. Previously there was a hypothesis,
formulated for the usual quantum Toda chain, that the product of
\emph{operators} $Q$, taken in the special points, is related to
the quantum Functional Bethe Ansatz. In this paper we have
established it explicitly, but for the product of \emph{modified
operators} $Q$. Below some explanations are given.

Let the solitonic amplitudes $g_k$ are in general position.  Let
$|\Psi_t^{(n)}\rangle$ be the complete set of eigenvectors of
$\tp^{(n)}(\lambda)$, and $\langle\Psi_t^{(n)}|$ be the
corresponding set of co-vectors,
\begin{equation}
\ds
\tp^{(n)}(\lambda)\;=\;\sum_t\;|\Psi_t^{(n)}\rangle\,t(\lambda)\,\langle\Psi_t^{(n)}|\;.
\end{equation}
Then
\begin{equation}\label{Q-expansion}
\ds
\Q^{(n)}_{\phi_n}\;=\;\sum_t\;|\Psi_t^{(n-1)}\rangle\,q_t(\lambda_n)\,\langle\Psi_t^{(n)}|\;,
\end{equation}
where $q_t(\lambda)$ is evidently the eigenvalues of simple
$Q$-operators, as in (\ref{BE}). Note, the functional Baxter
equation exists also and for arbitrary inhomogeneous chain, but to
obtain it from (\ref{BE-operator-12}) one has to use the
decomposition like (\ref{Q-expansion}).

With (\ref{Q-expansion}) taken into account, $\mathbf{K}$-operator
for general set of $g_k$ is
\begin{equation}\label{K-expansion}
\ds
\mathbf{K}\;=\;
\sum_t\;|\Psi_t^{(0)}\rangle\,q_t(\lambda_1)...q_t(\lambda_{M-1})\,\langle\Psi_t^{(M-1)}|\;,
\end{equation}
In the limit when all $g_k=1$, the sets $\langle\Psi^{(n)}_f|$
become degenerative, e.g.
\begin{equation}
\ds\langle\Psi^{(n)}_t|\,(\zop_1-\zop_M)\;=\;0\;\;\;\forall\;t\;.
\end{equation}
For us it is still mysterious what happens with the set
$|\Psi_t^{(n)}\rangle$ and co-set $\langle\Psi_t^{(n)}|$ when
$g_k=1$ -- this is the subject of separate investigation. But
nevertheless, in \cite{sp-1} and in this paper it is proven that
$\mathbf{K}$ is given by (\ref{K-expansion}) and simultaneously by
(\ref{K-separation}). Thus Sklyanin's formula appears explicitly:
\begin{equation}
\ds
\langle\Psi_t^{(0)}|\{\lambda_k\}_{k=1}^{M-1},\gamma\rangle\;=\;
\textrm{const}\;\prod_{k=1}^{M-1}\,q_t(\lambda_k)\;.
\end{equation}
Well known is that to solve Baxter's equation for $N>2$ is quite
hopeless (pure fermionic case $N=2$ is trivial, (\ref{BE}) may be
solved in one string). Only way is to avoid Baxter's equation for
high genus curve. So, we think, the attention is to be paid to the
special cases of the isospectral transfer matrices like $g_k=1$,
when several simplifications for the eigenstates are expected.

Note in conclusion, this method may be applied to any model, based
on the local Weyl algebra. Mentioned are to be the chiral Potts
model \cite{CPM,BS-cpm} and the Zamolodchikov-Bazhanov-Baxter
model in the vertex formulation \cite{mss-vertex}. To be exact,
all two dimensional integrable models with the local Weyl algebra
are particular cases of the general three dimensional model, and
their classical counterparts are known \cite{Sergeev}.

{\bf Acknowledgements} The authors are grateful to R. Baxter, V.
Bazhanov, V. Mangazeev, G. Pronko, E. Sklyanin, A. Belavin, Yu.
Stroganov, A. Isaev, G. von Gehlen, V. Tarasov, P. Kulish, F.
Smirnov and R. Kashaev for useful discussions and comments.

S.P. would also like to thank Max-Planck Instiut f\"ur Mathematik
(Bonn) for support and hospitality and S.S. would thank the
hospitality of MPIM during his short visit to Bonn supported by
Heisenberg-Landau program.

\bibliographystyle{amsplain}

\end{document}